\documentclass[aps,prl,twocolumn,showpacs,groupedaddress]{revtex4}
\usepackage{graphicx}
\usepackage{dcolumn}
\usepackage{bm}
\begin{document}
\title{Theoretical Analysis of Subthreshold Oscillatory Behaviors in Nonlinear Autonomous Systems}
\author{Shenbing Kuang}
\author{Jiafu Wang}
\email[]{jasper@whut.edu.cn}
\author{Ting Zeng}
\author{Aiyin Cao}
\affiliation{Department of Physical Science and Technology, Wuhan
University of Technology, Wuhan 430070 China}


\begin{abstract}
We have developed a linearization method to investigate the
subthreshold oscillatory behaviors in nonlinear autonomous systems.
By considering firstly the neuronal system as an example, we show
that this theoretical approach can predict quantitatively the
subthreshold oscillatory activities, including the damping
coefficients and the oscillatory frequencies which are in good
agreement with those observed in experiments. Then we generalize the
linearization method to an arbitrary autonomous nonlinear system.
The detailed extension of this theoretical approach is also
presented and further discussed.
\end{abstract}
 \maketitle

Oscillatory behaviors are one of the most important features for the
nonlinear systems and have attracted much attention in recent years
\cite{Chapman,BBK,Lefschetz}. Oscillation phenomena have been
discovered in many areas of the chemical and biological sciences,
such as chemical oscillation \cite{Kiss,Ikezoe}, rhythmic gene
expression and metabolism \cite{Schibler}, and particularly neuronal
oscillations \cite{Buzsaki2002,Buzsaki2004,Idoux}. These oscillatory
behaviors may have some observable influences on the responses of
the systems, and play significant and often crucial roles, for
example in neuronal system the oscillation activities are found to
be important for information processing \cite{Hutcheon,Schaefer} and
cognitive perception \cite{Ward,Ribary,Gelperin,Kaiser}. It is
therefore of great interest to study the oscillatory behaviors in
these nonlinear systems. Oscillation theory has become an important
part of contemporary applied mathematics \cite{Berezansky,Kulenovic}
and has been well developed \cite{EKZ,GL}. However, the analysis of
oscillatory processes is still far from being complete. Up to date,
little attention has been paid to the question how to systematically
characterize the oscillatory behaviors, especially regarding the
damping coefficients and oscillatory frequencies, in these nonlinear
systems.

Generally, the dynamics of nonlinear systems are often described by
some coupled differential equations \cite{EKZ,GL}. While the stimuli
to the system involve no variables that are differentiating
(constant stimuli for example) we call the nonlinear system an
autonomous one. In this Letter we attempt to approach the
aforementioned issue by presenting a linearization method to
theoretically investigate the subthreshold oscillation behaviors in
nonlinear autonomous systems. Firstly we consider the oscillatory
activities in the neuronal system under subthreshold constant
stimulus as an example. The theoretical approach enable us to obtain
analytical solutions for the damping coefficients and frequencies of
oscillatory behaviors of membrane voltage. In addition, we also
generalize the theoretical approach to an arbitrary nonlinear
autonomous system. The detailed implementation of this theoretical
framework is also presented and discussed.

Neuronal systems are highly nonlinear excitable systems. To
elucidate the subthreshold oscillatory behaviors in autonomous
neuronal system, we consider the Hindmarsh-Rose (HR) neural model,
as an example, under a subthreshold constant stimulus. The dynamics
of HR neuron system could be described by the following
equations\cite{HR1}:
\begin{equation}
\label{Eqs.1}
\begin{array}{lll}
        dx/dt = y - z - ax^{3} + bx^{2} + I(t), \\
        dy/dt = c - dx^{2} - y, \\
        dz/dt = r[s(x-x_{0}) - z],
\end{array}
\end{equation}
where $x$ denotes the membrane potential, $y$ the recovery variable
and $z$ a slow adaptation variable. and the parameters $a$, $b$,
$c$, $d$, $r$, $s$ and $x_0$ are the same as in Ref.\cite{HR1}. If
the membrane potential $x(t)$ exceeds its threshold value $x_{th} =
0$, the neuron will result in a spike $S(t) = \theta (x(t) -
x_{th})$, where $\theta (x)$ is the Heaviside function. $I(t)$
represents the stimulus bias to the neuron, $I(t) = I_0 < I_c$ with
$I_c$ being the critical value for constant stimulus.

It is well known that the HR neuron will be active with limit-circle
firings \cite{HR2} for a suprathreshold constant input, $I_{0}
> I_{c}$ in Eqs.\,(\ref{Eqs.1}). If $I_{0} < I_{c}$, the system may undergo a
temporal process of damping oscillation to reach a quiescent stable
state, with the membrane potential $x(t)$, the recovery variable
$y(t)$ and the slow adaptation variable $z(t)$ being the stable
value $x_s$, $y_s$, $z_s$, respectively. These stable values can be
obtained from Eqs.\,(\ref{Eqs.1}) by simply letting $dx/dt
 = dy/dt = dz/dt = 0$, yielding
\begin{equation}
\label{Eq.2}
  ax_{s}^{3}+(d-b)x_{s}^{2}+sx_{s}-(sx_{0} + c )=I_{0}
\end{equation}
and $y_s = c-dx_s^2$, $z_s=s(x_s-x_0)$. It is easy to prove that
Eq.\,(\ref{Eq.2}), with respect to $x_s$, has only one real root for
any value of $I_0$,
\begin{equation}
\label{Eq.3}
x_{s}= (-Q/2+\sqrt{\Delta})^{1/3}
+(-Q/2-\sqrt{\Delta})^{1/3}-(d-b)/3a
\end{equation}
where $\Delta = (Q/2)^{2} + [s/3a-(d-b)^{2}/9a^{2}]^{3}$ and
      $Q = 2(d-b)^{3}/27a^{3} - s(d-b)/3a^{2} - (sx_{0} + c + I_{0})/a$.
One can always write
\begin{equation}
\label{Eqs.4}
\begin{array}{lll}
   x(t) = x_{s} + \tilde{x}(t),\\
   y(t) = y_{s} + \tilde{y}(t),\\
   z(t) = z_{s} + \tilde{z}(t),
\end{array}
\end{equation}
Since $I_{0} < I_{c0}$, the variables $\tilde{x}(t)$, $\tilde{y}(t)$
and $\tilde{z}(t)$ are small (approximately zero) for sufficiently
large $t$; so one can linearize Eqs.\,(\ref{Eqs.1}), by utilizing
Eqs.\,(\ref{Eqs.4}) into the following equation
\begin{equation}
\label{Eqs.5}
  \left[
    \begin{array}{lll}
    d\tilde{x}/dt \\
    d\tilde{y}/dt \\
    d\tilde{z}/dt
    \end{array}
  \right]
  =
  \left[
    \begin{array}{ccc}
    2bx_{s}-3ax_{s}^{2} &  \;  1 & \;  -1  \\
      -2dx_{s}          &  \; -1 & \;   0  \\
         rs             &  \;  0 & \;  -r
    \end{array}
  \right]
  \left[
    \begin{array}{lll}
    \tilde{x} \\
    \tilde{y} \\
    \tilde{z}
    \end{array}
  \right]
\end{equation}
The eigenvalue $\lambda$ of the matrix in Eqs.\,(\ref{Eqs.5}) yields
\begin{equation}
\label{Eq.6}
 (\lambda+r)[(\lambda+1)(\lambda+3ax_{s}^{2}-2bx_{s})+2dx_{s}]
   + rs(\lambda + 1) = 0
\end{equation}
When $0 \leq I_{0} < I_{c}$, Eq.\,(\ref{Eq.6}) has one real root,
which remains negative, and two conjugated complex roots, say
$\lambda_{r} \pm i\lambda_{i}$. These roots, as well as $x_{s}$,
depend on $I_{0}$ and can be obtained analytically through
Eqs.\,(\ref{Eq.3}) and (\ref{Eq.6}):
\begin{equation}
\label{Eqs.7}
\begin{array}{lll}
\lambda_{r}(I_{0}) =
-[(-\epsilon/2 + \sqrt{\delta})^{1/3} + (-\epsilon/2 - \sqrt{\delta})^{1/3}]/2 - \alpha /3, \\
\lambda_{i}(I_{0}) =
\sqrt{3}[(-\epsilon/2-\sqrt{\delta})^{1/3}-(-\epsilon/2+\sqrt{\delta})^{1/3}]/2,\\
\end{array}
\end{equation}
where $\delta = (\epsilon/2)^{2} + [(-\alpha^{2}/3 + \beta)/3]^{3}$,
    $\epsilon = 2\alpha^{3}/27 - \alpha \beta/3 + \gamma$,
    $\alpha = 3ax_{s}^{2}-2bx_{s}+r+1$,
    $\beta = 3a(1+r)x_{s}^{2}+2(d-b)x_{s}+r(s+1)$ and
    $\gamma = r[3ax_{s}^{2}+2(d-b)x_{s}+s]$ with $x_s$ obtained from Eq.\,(\ref{Eq.3}).

The analytical results for the dependence of $ \lambda_{r} $ upon
$I_0$ are shown in figure 1. When $I_{0}$ increases from $0$ to the
threshold value $I_{c}$, the negative $ \lambda_{r} $ increases
monotonically up to the critical value $0$ corresponding to the
limit-circle firings behaviors observed in simulation \cite{HR2}. Be
aware that there still exists damped behaviors even when the input
to the neuron is a inhibitory one. The imaginary part
    $\lambda_{i} \ne 0$
means that the membrane potential $x(t)$, as a function of the real
time $t$, behaves a damping oscillation with an intrinsic frequency
$f_{i} \equiv \lambda_{i}$. The calculation results of $f_{i}$
versus $I_{0}$ are shown in figure 2. One can see that the intrinsic
frequency $f_{i}$ of the damping oscillation is about $10 \sim 40$
Hz, depending on $I_{0}$. It is notable that the damping oscillation
does exist even when $I_0 = 0$, indicating the inherence of the
oscillation. Obviously, this frequency range is the same as the
subthreshold activity observed in experiments
\cite{Llinas,Lampl,Buzsaki2002,Mann,Bartos}.

In fact, our theoretical analysis of subthreshold oscillation
activities in neuronal systems are naturally independent of the
specific neuronal model used. The intrinsic oscillatory behaviors in
other excitable neuronal models, such as the Bonhoeffer van der Pol
(BvP) model, FitzHugh-Nagumo (FHN) model and Hodgkin-Huxley -type
(HH) one, can also be obtained analytically or numerically provided
that the stimuli to these nonlinear systems are subthreshold
constant ones. In the following we will describe a generic version
of our theoretical approach, for the case of an arbitrary nonlinear
autonomous system under subthreshold stimuli, to investigate the
intrinsic oscillatory behaviors with an emphasis on the damping
coefficients and oscillatory frequencies.

The dynamics of nonlinear autonomous systems are generally described
by the differential equations of the first order or higher order for
complex systems, and the latter can easily be reduced to several
coupling differential equations of first order. In this study, we
suppose that an autonomous nonlinear system is governed by the
following $n$ coupling equations:
\begin{equation}
\label{Eqs.8}
 dx_i/dt = f_i(x_1, x_2, ... x_n) + A_i     \; (i = 1, 2, ...n)
\end{equation}
where $A_i$ are the constant stimuli to the system and make the
nonlinear system an autonomous one. We simply let $dx_i/dt = 0$ and
easily get the following $n$ equations with real coefficients:
\begin{equation}
\label{Eqs.9}
 f_i(x_1,x_2, ... x_n) = 0,  \; (i = 1, 2, ... n)
\end{equation}
In principle, there are up to $n$ real roots for the above
equations, indicating that the nonlinear autonomous system has $n$
steady states at most. Generally, the Eqs.\,(\ref{Eqs.9}) may have
both real roots and complex roots, the former of which implies the
number of steady states while the latter should appear in pairs of
conjugated complex due to the real coefficients of
Eqs.\,(\ref{Eqs.9}). If all the roots of the Eqs.\,(\ref{Eqs.9}) are
complex ones, we can conclude that the dynamic of the nonlinear
autonomous system will behavior divergently or chaotically and the
nonlinear system is no longer convergent to any steady states, which
is beyond our consideration in this study.

For simplicity let us pay close attention to one of the steady state
of system, $x_s$, where each variables of the system reaches their
own steady value, $x_{is}$. During the convergent process of the
system one can always write
\begin{equation}
\label{Eq.10}
 x_i = x_{is} + \tilde x_i,           \; (i = 1, 2, ... n)
\end{equation}
whereas $\tilde x_i$ are small enough for sufficiently long time
$t$. Inserting Eq.\,(\ref{Eq.10}) into Eqs.\,(\ref{Eqs.8}) we can
get
\begin{equation}
\label{Eqs.11}
d\tilde x_i/dt = f_i(x_{1s}+\tilde x_1, x_{2s}+\tilde
x_2, ... x_{ns}+\tilde x_n)  \; (i =1, 2, ... n)
\end{equation}
where the expression terms on the right side of Eqs.\,(\ref{Eqs.11}
can be further expanded at the steady state, $x_s$, up to the
accuracy of first order Taylor expansion since $\tilde x_i$ are
sufficient small variables, into the following as:
\begin{equation}
\label{Eqs.12}
\begin{array}{lcr}
  $ $ f_i(x_{1s}+\tilde x_1, x_{2s}+\tilde x_2, ... x_{ns}+\tilde x_n) \\
  $ = $ f_i(x_{1s}, x_{2s}, ... x_{ns}) \\
  $ + $ \tilde x_{1}\frac{\partial{f_i}}{\partial{x_1}}|_{x_s} + \tilde x_{2}\frac{\partial{f_i}}{\partial{x_2}}|_{x_s}
  $ + $ ... + \tilde x_{n} \frac {\partial{f_i}} {\partial{x_n}}| _{x_s},     \; (i = 1, 2, ... n)
\end{array}
\end{equation}
After inserting Eqs.\,(\ref{Eqs.9}) and Eqs.\,(\ref{Eqs.12}) into
Eqs.\,(\ref{Eqs.11}) we obtain $n$ new differential equations
expressed below in the form of the matrix.
\begin{equation}
\label{Eqs.13}
  \left[
    \begin{array}{ccccc}
    d\tilde {x_1}/dt \\
     ...          \\
    d\tilde {x_i}/dt \\
     ...          \\
    d\tilde {x_n}/dt
    \end{array}
  \right]
  =
  \left[
    \begin{array}{ccccc}
 \frac{\partial{f_1}} {\partial{x_1}}| _{x_s}   &  \;  ...  &  \;  \frac{\partial{f_1}} {\partial{x_i}}| _{x_s}
            &  \; ... &  \;  \frac{\partial{f_1}} {\partial{x_n}}| _{x_s}                         \\
     ...            &  \;  ...   &  \;      ...           & \;  ...  & \;     ...                 \\
 \frac{\partial{f_i}} {\partial{x_1}}| _{x_s}   &  \;  ...  &  \;  \frac{\partial{f_i}} {\partial{x_i}}| _{x_s}
            &  \; ... &  \;  \frac{\partial{f_i}} {\partial{x_n}}| _{x_s}                         \\
     ...            &  \;  ...   &  \;      ...           & \;  ...  & \;     ...                 \\
 \frac{\partial{f_n}} {\partial{x_1}}| _{x_s}   &  \;  ...  &  \;  \frac{\partial{f_n}} {\partial{x_i}}| _{x_s}
            &  \; ... &  \;  \frac{\partial{f_n}} {\partial{x_n}}| _{x_s}
    \end{array}
  \right]
  \left[
    \begin{array}{ccccc}
    \tilde{x_1} \\
    ...     \\
    \tilde{x_i} \\
    ...     \\
    \tilde{x_n}
    \end{array}
  \right]
 \end{equation}
The eigenvalues of the $n\times n$ real matrix in
Eqs.\,(\ref{Eqs.13}) can thus be expressed as a function of steady
values $x_s$. Since the coefficients of the matrix in
Eqs.\,(\ref{Eqs.13}) are real, the coefficients of the eigenfunction
should also be real numbers. In general the eigenvalues of the
matrix with real coefficients should be either real numbers or in
pairs of conjugated complex. For the case of real numbers, the
eigenvalues would remain negative, corresponding to the convergence
of the nonlinear system under subthreshold stimuli. Otherwise the
system may undergo divergent or chaotic dynamics; For the case of
pairs of conjugated complex, the real part of eigenvalue must be
negative suggesting the damping factors of the system, while the
imaginary part implies the frequencies of the oscillatory
activities. As a rule, there are up to [$\frac{n}{2}$] components of
oscillatory frequencies in this nonlinear autonomous system ([]
means the biggest integer that no greater than what's in the square
bracket).

At the any other steady state beyond $x_s$, the same linearizaiton
method could also be applied. The key of our theoretical approach is
that the stimuli to the nonlinear autonomous system must be
subthreshold ones, to keep the nonlinear system from divergence or
chaos, so that the linearization method is always available.

In summary, we have developed a linearization method to study the
subthreshold oscillatory behaviors in nonlinear autonomous systems.
This theoretical approach enable us to predict the dynamics of
subthreshold behaviors and quantitatively describe the damping
coefficients and frequencies of oscillatory behaviors, if exist.
Given the case of nonlinear neuronal system under subthreshold
constant stimuli, our theoretical analysis can give analytical
solutions about its oscillation behaviors that are consistent with
experimental observations. Since oscillatory activities are
surprisingly ubiquitous phenomena in a variety of nonlinear systems,
the linearization method provide us a possible theoretical approach
to characterize the oscillatory behaviors in those autonomous
systems. Regarding the oscillatory behaviors in the non-autonomous
nonlinear system, one have to seek other methods or solutions which
are our interest of further study.

\section{Acknowledgements}
\begin{acknowledgments}
The authors would like to acknowledge the financial support from the
Key Project of Chinese Ministry of Education (Grant No. 106115).
\end{acknowledgments}

\bibliographystyle{unsrt}
\bibliography{NonlinearOsc}

\begin{figure}
\includegraphics{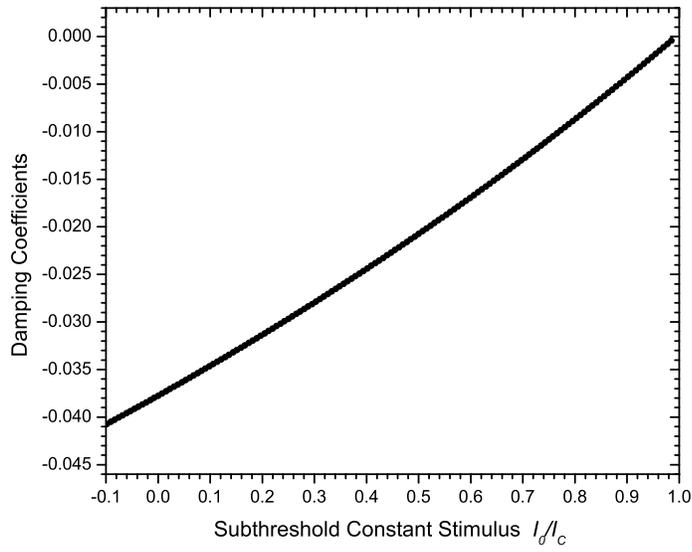}
\caption{\label{figure1} Theoretical prediction of damping
coefficients versus constant stimuli $I_0$. $I_c$ is estimated
numerically as the critical value of stimulus that make the membrane
potential exceed its threshold value $x_{th} = 0$ after a sufficient
long time, and chosen as 1.32.}
\end{figure}

\begin{figure}
\includegraphics{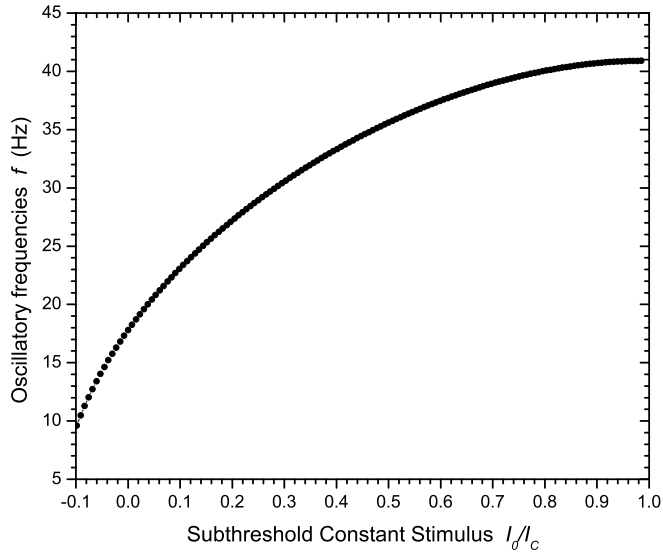}
\caption{\label{figure2} Theoretical prediction of oscillatory
frequencies versus constant stimuli $I_0$. $I_c$ is chosen as the
same as that in figure 1.}
\end{figure}

\end{document}